\documentclass[prb,preprint,floatfix,superscriptaddress]{revtex4-1}

\usepackage{epsf}
\usepackage{epsfig}
\usepackage{graphicx}
\usepackage{dcolumn}
\usepackage{bm}
\usepackage{amsfonts}
\usepackage{amsmath}

\usepackage{amssymb}
\usepackage{color,soul}

\input epsf

\begin{document}
\title{Single-parameter spin-pumping in driven metallic rings with spin-orbit coupling}
\author{J. P. Ramos}
\affiliation{Departamento de F\'{\i}sica, Universidad Cat\'{o}lica del Norte, Angamos 0610, Casilla 1280, Antofagasta, Chile}
\author{L. E. F. Foa Torres}
\affiliation{Instituto de F\'{\i}sica Enrique Gaviola (CONICET) and FaMAF, Universidad Nacional de C\'ordoba, Ciudad Universitaria 5000, C\'{o}rdoba, Argentina}
\author{P. A. Orellana}
\affiliation{Departamento de F\'{\i}sica, Universidad Federico Santa Mar\'ia, Avenida Vicu\~{n}a Mackenna 3939, San Joaquin, Santiago, Chile}
\author{V. M. Apel}
\affiliation{Departamento de F\'{\i}sica, Universidad Cat\'{o}lica del Norte, Angamos 0610,  Casilla 1280, Antofagasta, Chile}

\begin{abstract}
We consider the generation of a pure spin-current at zero bias voltage with a single time-dependent potential. To such end we study a device made of a mesoscopic ring connected to electrodes and clarify the interplay between a magnetic flux, spin-orbit coupling and non-adiabatic driving in the  production of a spin and electrical current. By using Floquet theory, we show that the generated spin to charge current ratio can be controlled by tuning the spin-orbit coupling.
\end{abstract}
\date{\today}
\maketitle

\section{Introduction}
Largely forgotten during the early decades of nanoelectronics, the spin degree of freedom is becoming ever closer to the center of the research stage \cite{Wolf16112001,pulizzi2012}. Indeed, generating and detecting spin-currents is now a fascinating field of research \cite{Zutic2004,Sinova2012} with applications in future electronics \cite{Jansen2012}, quantum computing \cite{Awschalom08032013} and information storage \cite{McCamey17122010}. Among the many ways of harnessing the electron spin, spin orbit interaction (SOI) in two-dimensional electron gases is a promising one since the spin transport properties can be controlled simply by applying an electric field \cite{datta,nitta}.

Most of the proposals aiming at the control of the spin degree of freedom use static electric or magnetic fields \cite{Zutic2004,Sinova2012}. Here we follow a different path and use alternating fields (ac) as in \cite{Citro2006,Wu2007}. The time-dependence introduced by the alternating fields \cite{Kohler2005} provide an avenue for exploring new phenomena including the opening of a laser-induced bandgap \cite{Wang2013,Calvo2011} or chiral edge states \cite{Perez-Piskunow2013} and, more generally, the tuning of its topological properties \cite{Lindner2011,Perez-Piskunow2013,Katan2013,Sen2013}. Another striking phenomena is the coherent generation of a current at zero bias voltage (termed \textit{quantum charge pumping}) \cite{Thouless1983,Buettiker2006,Moskalets2002} and, as shown below, the generation of a pure spin-currents through spin pumping. Quantum pumping is usually achieved by driving a sample connected to electrodes through ac gate voltages. Within the adiabatic approximation \cite{Brouwer1998}, pumping a non-vanishing charge requires the presence of at least two time-dependent parameters (typically constituted by gate voltages) and has been widely studied in many systems including pristine \cite{Prada2009,Zhu2009} and disordered graphene \cite{Ingaramo2013}. But beyond the adiabatic approximation \textit{single-parameter pumping} is also possible as predicted theoretically \cite{Shutenko2000,Arrachea2005,FoaTorres2005} (similar to the mesoscopic photovoltaic effect predicted earlier in \cite{Falko1989}) and achieved in careful experiments \cite{Kaestner2008,Fujiwara2008,Kaestner2009}. Besides reducing the burden of adding more contacts in a nanoscale sample, a single parameter setup could also prove advantageous (as a compared to a two-parameter one) in reducing capacitive effects and crosstalk between time-dependent gates.

\begin{figure}[htpb]
\includegraphics[width=0.5\columnwidth]{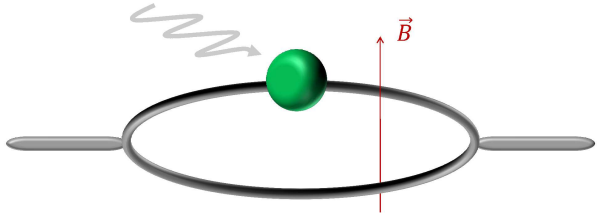}
\caption{Scheme of setup considered in the text, a quantum ring with a magnetic field cross them and a quantum dot embedded in one of its arms, driven by an ac voltage source.}
\label{fig1}
\end{figure}

Here we address the effect of spin-orbit coupling and its interplay with a single time-dependent field in the generation of non-adiabatic spin current at zero bias voltage. To such end we consider a setup as the one represented in Fig.\ref{fig1}, where a nanoscale ring is connected to electrodes and has a quantum dot embeded in one of its arms. The time dependence is introduced as an alternating gate applied to the quantum dot and does not break neither time-reversal nor inversion symmetry (parity). Crucial to the generation of pumped current is the addition of a magnetic flux threading the ring as shown in Fig.~(\ref{fig1}). The spin-orbit coupling is introduced as an additional spin-dependent flux.
In this paper we show how this simple setup is able to provide a minimal model where  a pure spin-current can be achieved.

\section{Hamiltonian Model and its solution through Floquet Theory}

Let us start our discussion by presenting our model Hamiltonian for the situation represented in Fig.(\ref{fig1}). The total Hamiltonian $\cal{H}(t)$ is written as:
\begin{equation}
{\cal H}(t)={\cal H}_{C}+{\cal H}_{QD}(t)+{\cal H}_{T},
\end{equation}
where ${\cal H}_{C}$ represents the left and right contacts and the lower arm of the ring (represented by site $j=0$ in the notation below), ${\cal H}_{QD}(t)$ the quantum dot in the upper arm of the ring (which for simplicity is taken to be a single level), and ${\cal H}_{T}$ the tunneling Hamiltonian between the quantum-ring and the contacts, which are given by
\begin{eqnarray}
&&{\cal H}_{C}=\sum_{j=-\infty,\sigma}^{\infty}(\varepsilon_{j}c_{j,\sigma}^{\dag}c_{j,\sigma}+\gamma c_{j,\sigma}^{\dag}c_{j+1,\sigma})+\text{h.c.},\\
&&{\cal H}_{QD}(t)=\varepsilon_{d}(t)\sum_{\sigma}d_{\sigma}^{\dag}d_{\sigma}\\
&&{\cal H}_{T}=\sum_{\sigma}(V_{L}^{\sigma}c_{-1,\sigma}^{\dag}d_{\sigma}+V_{R}c_{1,\sigma}^{\dag}d_{\sigma})+\text{h.c.}.
\end{eqnarray}
The time dependence is introduced as a modulation of the energy levels of the quantum dot. For a single-level quantum-dot, this is is achieved through $\varepsilon_0(t)=\varepsilon_{0}+v\cos(\Omega_{0}t)$. We consider a magnetic and electric fields in the system, their contributions to the Hamiltonian are embedded in the hopping matrix elements $V_{L}^{\sigma}=V_{0}\exp[i2\pi(\phi_{\text{AB}}+\sigma\phi_{\text{SO}})/\phi_{0}]$, where $\phi_{\text{AB}}$ and $\phi_{\text{SO}}$ are the phases due to the Aharonov-Bohm effect and spin-orbit interaction respectively, $\sigma$ is the spin index ($\sigma=\,\uparrow,\downarrow$ or $\sigma=1,-1$) and $\phi_{0}$ is the flux quantum.

Since we are interested in a single-parameter pumping configuration as in \cite{Kaestner2008,FoaTorres2005,Agarwal2007,FoaTorres2011}, the calculation of the electrical response requires going beyond the adiabatic theory. Floquet theory offers a suitable framework \cite{Moskalets2002,Kohler2005}. Here we use it in combination with Green's functions, then we have a Floquet-Green function denoted by $G_{F}$ defined from the Floquet's Hamiltonian ${\cal  H_{F}}$ as \cite{FoaTorres2005,FoaTorres2009} $G_{F}=[E{\cal I}-{\cal H_{F}}]^{-1}$.

If the spin-orbit coupling does not couple different spin channels, as in our case, the dc component of the current is given by:

\begin{eqnarray}\label{Floquet-Current}
&&\bar{I}_{\sigma}=\frac{1}{\tau}\int_0^\tau dt I_{\sigma}(t) \\
&&\bar{I}_{\sigma}=\frac{e}{h}\times\\
&&\sum_{n}\int\left[T_{(R,\sigma),(L,\sigma)}^{(n)}(\varepsilon)f_{L}(\varepsilon)-T_{(L,\sigma),(R,\sigma)}^{(n)}(\varepsilon)f_{R}(\varepsilon)\right]d\varepsilon,\nonumber
\end{eqnarray}
where $T_{(R,\sigma),(L,\sigma)}^{(n)}(\varepsilon)$ is the probability for an electron on the left ($L$) with spin $\sigma$ and energy $\varepsilon$  to be transmitted to the right ($R$) reservoir while exchanging $n$ photons and $\tau=2\pi/\Omega_0$. These probabilities are weighted by the usual Fermi-Dirac distribution functions $f_{R(L)}$ for each electrode and are given, in terms of Floquet-Green function, by
\begin{equation}
\label{transmision}
T_{(R,\sigma),(L,\sigma)}^{(n)}(\varepsilon)=4\Gamma_{R}(\varepsilon+n\hbar\Omega_{0})|G_{LR,\sigma}^{(n)}(\varepsilon)|^{2}\Gamma_{L}(\varepsilon)\,,
\end{equation}
where the probability in opposite direction is described exchanging the index $L$ with $R$, and $\Gamma_{L(R)}$ is the matrix coupling with left (right) electrode, defined as the imaginary part of the electrode's self energies, i. e. $\Gamma_{L(R)}=-\text{Im}(\Sigma_{L(R)})$.

The associated spin-current is $\bar{I}_{\text{s}}=\bar{I}_{\uparrow}-\bar{I}_{\downarrow}$ while the charge current is $\bar{I}=\bar{I}_{\uparrow}+\bar{I}_{\downarrow}$.

\section{Results and discussion}

Using the model introduced before we now turn to our results for the pumped electric and spin currents. To start with we consider the system in the absence of spin-orbit coupling. We consider the leads in thermodynamic equilibrium (\textit{i. e.} $f_{L}(\varepsilon)=f_{R}(\varepsilon)=f(\varepsilon)$) as a semi-infinite 1d system with nearest neighbor coupling $\gamma$, which is used as energy parameter. The ac field frequency is set to $\Omega_{0}$ such that $\hbar\Omega_{0}=\gamma/5$, and the field magnitude is $v=0.07\,\gamma/e$. The hopping between the contacts and QD is $V_{0}=\gamma/4$.

\begin{figure}[htpb]
\includegraphics[width=8cm]{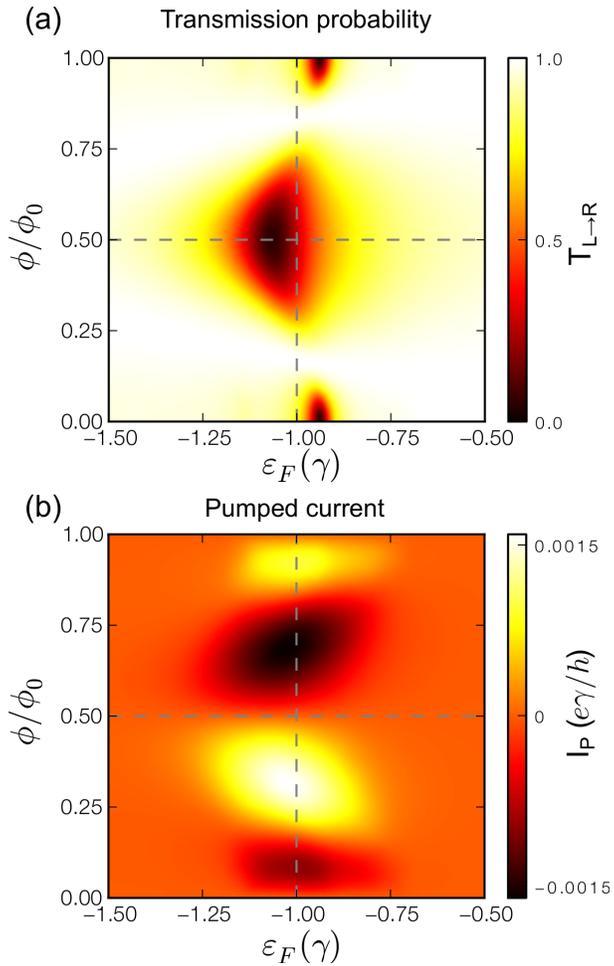}
\caption{(Color online) (a) Transmission probability from left to right as a function of the applied magnetic flux and the Fermi energy (for vanishing spin-orbit interaction). (b) Same as (a) for the pumped current. Note the emergence of local maxima/minima close to the parameters where a transmission zero is observed.}
\label{fig2}
\end{figure}

Figure \ref{fig2}a shows a contour plot of the transmission probability as a function of the Fermi level position and the magnetic flux. There we can observe the presence of a region where the transmission is very close to zero (close to the intersection of the dashed lines). This is due to a destructive quantum interference known as Fano resonance or antiresonance \cite{Guinea1987,DAmato1989,LadrondeGuevara2006}(for a recent review \cite{Miroshnichenko2010}). Interestingly, the pumped current shown in Fig. \ref{fig2}b achieves a maximum intensity whenever the parameters are tuned close to the transmission zero. Besides, we can see that the sign of the observed maxima is reversed when traversing the transmission zero. 

We note that a single-time dependent harmonic potential does not break time reversal symmetry (being defined as the existence of a time $t_0$ such that the Hamiltonian which is a function of the time $t$ satisfies ${\cal H}(t_0+t)={\cal H}(t_0-t)$). It is the magnetic field that breaks TRS and allows for pumping to occur. Note, however that this is true only whenever magnetic flux is different from the half integer multiples of the flux quantum. For a magnetic flux of $\pi$ for example, the Hamiltonian does not change upon time-reversal (the phase in the hopping term $V_L$ changes from $\exp(i\pi)$ to $\exp(-i\pi)$ and therefore there is no pumped current as observed in Fig. 2-b. On the other hand one should note that the magnetic field alone would not produce pumping. This is the point where the time-dependent field enters into the game. Its role in this setup is to provide for additional effective channels for transport, thereby circumventing the constraint of phase-rigidity \cite{Buettiker1988} and allowing for the directional asymmetry in the transmission probabilities. 

The addition of spin-orbit interaction breaks the spin degeneracy and at zero bias both charge \textit{and} spin currents are generated. By examining Fig. (\ref{fig2}) one can imagine that the spin-orbit phase may be used to tune the working point of our pump for each spin \textit{independently}. Indeed, the term $\phi_{\text{AB}}+\sigma\phi_{\text{SO}}$ enters as an effective spin-dependent flux $\phi_{\sigma}^{\text{eff}}$ . In particular, we could choose this spin-orbit phase so that it cancels out for one spin direction (leading to a vanishing pumped charge for this spin) and adds up for the other, or in such a way as to cancel the charge current while summing up towards the spin current.

\begin{figure}[htpb]
\includegraphics[width=0.7\columnwidth]{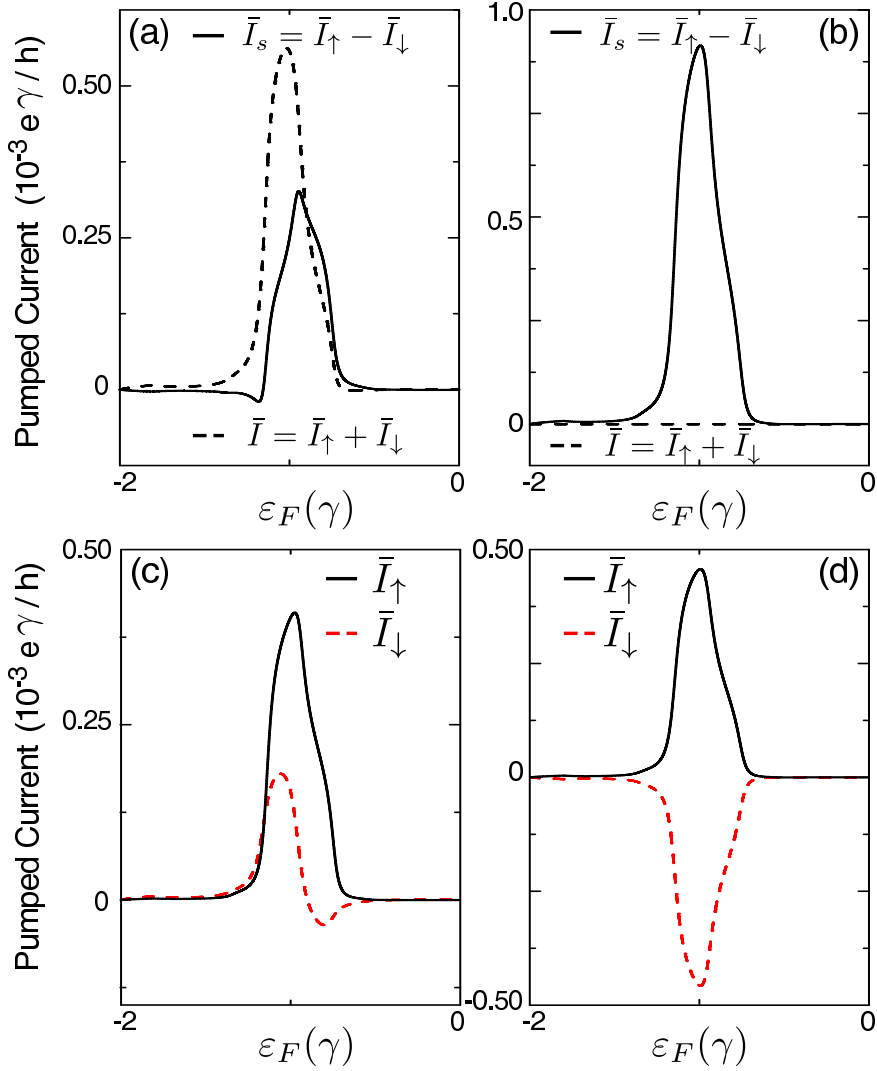}
\caption{(Color online) a-b Pumped charge ($\bar{I}$, dashed line) and spin currents ($\bar{I}_{s}$, solid line), dashed and solid for different values of the applied static flux and spin-orbit interaction: (a) $\phi_{AB}=0.2\phi_{0}$, $\phi_{SO}=0.1\phi_{0}$ and b) $\phi_{AB}=0.5\phi_{0}$, $\phi_{SO}=0.4\phi_{0}$. One can see that in (b) the charge current vanishes but the spin current is enhanced. The spin-resolved contributions to the current for the same cases are shown in (c) and (d).}
\label{fig3}
\end{figure}

Figures \ref{fig3} (a) and (b) show the charge (solid lines) and spin (dashed lines) currents as a function of the Fermi energy for different values of the spin-orbit and Aharanov-Bohm phases, while Figures \ref{fig3} (c) and (d) show the pumped current for each spin, spin up (black solid lines) and down (red dashed lines). As anticipated, the parameters can be chosen so that the currents for each spin direction have opposite signs (Fig. \ref{fig3} (d)), thereby leading to a pure spin-current as on Fig.\ref{fig3} (b). In this situation, the charge current cancels out whether the spin-current is maximal.

A point that needs to be emphasized in this proposal is that the pumped current is intrinsically non-adiabatic (this contrasts for example with Ref. \cite{Citro2006} using a similar setup but with two time-dependent parameters). An adiabatic calculation would actually give a vanishing response. Going beyond this adiabatic (low-frequency) limit is therefore mandatory justifying the use of Floquet theory. On the other hand, the pumped currents in this case emerges as an interplay between photon-assisted processes and the interference in the Aharanov-Bohm ring \cite{FoaTorres2005}. A similar setup but without contacts to electrodes were considered in \cite{Kravtsov1993,Yudson2003}. The spin-orbit coupling allows to obtain spin polarized pumped currents and the key role of the time-dependent field is to provide for additional paths for interference breaking phase-rigidity \cite{Buettiker1988}, although it  does not break time-reversal symmetry.

The setup discussed here can be realized by using the present technologies. A quantum dot inserted in a mesoscopic ring has been fabricated by several laboratories in the last decades\cite{Buks1996,Kobayashi2003}. A particularly interesting case would be an $InGAs$ quantum-dot inserted in a mesoscopic quantum-ring since $InGAs$ has a strong spin-orbit coupling and this coupling can be controlled by an electric field \cite{Amaha2008}.

\section{Final remarks}
In summary, we study quantum spin-pumping with a single parameter in a configuration where the effect of the time-dependent field is reduced to the essential one: providing for additional channels for transport. The spin-orbit interaction breaks the spin degeneracy and we exploit it to generate a pure pumping spin-current through the independent tuning of the phases dues to Aharonov-Bohm effect and spin-orbit coupling.

\section{Acknowledgments}
We acknowledge support of the SPU-Argentina (Project PPCP-025), SeCyT-UNC, ANPCyT-FonCyT, FONDECYT project number 1100560, CONICYT (Project PSD-65), and the scholarship CONICYT academic year 2012 number 22121816.\\


\begin{thebibliography}{10}

\bibitem{Wolf16112001}
S.~A. Wolf {\it et~al.}, Science {\bf 294},  1488  (2001).

\bibitem{pulizzi2012}
F. Pulizzi, Nature {\bf 11},  367  (2012).

\bibitem{Zutic2004}
I. Zutic, I., J. Fabian, and S. Das~Sarma, Rev. Mod. Phys. {\bf 76},  323
  (2004).

\bibitem{Sinova2012}
J. Sinova and I. Zutic, Nat Mater {\bf 11},  368  (2012).

\bibitem{Jansen2012}
R. Jansen, Nature Materials {\bf 11},  400  (2012).

\bibitem{Awschalom08032013}
D.~D. Awschalom {\it et~al.}, Science {\bf 339},  1174  (2013).

\bibitem{McCamey17122010}
D.~R. McCamey, J. Van~Tol, G.~W. Morley, and C. Boehme, Science {\bf 330},
  1652  (2010).

\bibitem{datta}
S. Datta and B. Das, Applied Physics Letters {\bf 56},  665  (1990).

\bibitem{nitta}
J. Nitta, F.~E. Meijer, and H. Takayanagi, Applied Physics Letters {\bf 75},
  695  (1999).

\bibitem{Citro2006}
R. Citro and F. Romeo, Phys. Rev. B {\bf 73},  233304  (2006).

\bibitem{Wu2007}
B.~H. Wu and J.~C. Cao, Phys. Rev. B {\bf 75},  113303  (2007).

\bibitem{Kohler2005}
S. Kohler, J. Lehmann, and P. H\"{a}nggi, Physics Reports {\bf 406},  379
  (2005).

\bibitem {Wang2013} Y. H. Wang, H. Steinberg, P. Jarillo-Herrero, N. Gedik, Science \textbf{342}, 453 (2013).

\bibitem {Calvo2011} T. Oka and H. Aoki, Phys. Rev. B {\bf 79}, 081406 (2009); H. L. Calvo, H. M. Pastawski, S. Roche, L. E. F. Foa Torres, Appl. Phys. Lett. {\bf 98}, 232103 (2011).

\bibitem {Perez-Piskunow2013} P. M. Perez-Piskunow, G. Usaj, C. A. Balseiro, L. E. F. Foa Torres, Phys. Rev. B {\bf 89}, 121401(R) (2014).

\bibitem{Lindner2011} N. H. Lindner, G. Refael, V. Galitski, Nature Physics \textbf{7}, 490 (2011).


\bibitem {Katan2013} Y. Tenenbaum Katan and D. Podolsky, Phys. Rev. B {\bf 88}, 224106 (2013).

\bibitem {Sen2013} M. Thakurathi, K. Sengupta, D. Sen, Majorana edge modes in the Kitaev model, arxiv:
arXiv:1310.4701 [cond-mat.mes-hall] (unpublished).

\bibitem{Thouless1983}
D.~J. Thouless, Phys. Rev. B {\bf 27},  6083  (1983).

\bibitem{Buettiker2006}
M. B\"{u}ttiker and M. Moskalets,  in {\em Lecture Notes in Physics}, edited by
  J. Asch and A. Joye (Springer Berlin Heidelberg , 2006), Vol.~690, pp.\
  33--44--.

\bibitem{Moskalets2002}
M. Moskalets and M. B\"{u}ttiker, Phys. Rev. B {\bf 66},  205320  (2002).

\bibitem{Brouwer1998}
P.~W. Brouwer, Phys. Rev. B {\bf 58},  R10135  (1998).

\bibitem{Prada2009}
E. Prada, P. San-Jose, and H. Schomerus, Phys. Rev. B {\bf 80},  245414
  (2009).

\bibitem{Zhu2009}
R. Zhu and H. Chen, Applied Physics Letters {\bf 95},  122111  (2009).

\bibitem{Ingaramo2013}
L. Ingaramo and L.~E.~F. Foa~Torres, Appl. Phys. Lett. {\bf 103}, 123508 (2013).

\bibitem{Shutenko2000}
T.~A. Shutenko, I.~L. Aleiner, and B.~L. Altshuler, Phys. Rev. B {\bf 61},
  10366  (2000).

\bibitem{Arrachea2005}
L. Arrachea, Phys. Rev. B {\bf 72},  121306  (2005).

\bibitem{FoaTorres2005}
L.~E.~F. Foa~Torres, Phys. Rev. B {\bf 72},  245339  (2005).

\bibitem{Falko1989}
V. Fal’ko and D. Khmelnitskii, Sov. Phys. JETP {\bf 68},  186  (1989).

\bibitem{Kaestner2008}
B. Kaestner {\it et~al.}, Phys. Rev. B {\bf 77},  153301  (2008).

\bibitem{Fujiwara2008}
A. Fujiwara, K. Nishiguchi, and Y. Ono, Appl. Phys. Lett. {\bf 92},  042102
  (2008).

\bibitem{Kaestner2009}
B. Kaestner {\it et~al.}, Appl. Phys. Lett. {\bf 94},  012106  (2009).

\bibitem{Agarwal2007}
A. Agarwal and D. Sen, Phys. Rev. B {\bf 76},  235316  (2007).

\bibitem{FoaTorres2011}
L.~E.~F. Foa~Torres, H.~L. Calvo, C.~G. Rocha, and G. Cuniberti, Appl. Phys.
  Lett. {\bf 99},  092102  (2011).

\bibitem{FoaTorres2009}
L.~E.~F. Foa~Torres and G. Cuniberti, Appl. Phys. Lett. {\bf 94},  222103
  (2009).

\bibitem{Guinea1987}
F. Guinea and J.~A. Verg\'{e}s, Phys. Rev. B {\bf 35},  979  (1987).

\bibitem{DAmato1989}
J.~L. D'Amato, H.~M. Pastawski, and J.~F. Weisz, Phys. Rev. B {\bf 39},  3554
  (1989).

\bibitem{LadrondeGuevara2006}
M.~L. Ladron~de Guevara and P.~A. Orellana, Phys. Rev. B {\bf 73},  205303
  (2006).

\bibitem{Miroshnichenko2010}
A.~E. Miroshnichenko, S. Flach, and Y.~S. Kivshar, Rev. Mod. Phys. {\bf 82},
  2257  (2010).

\bibitem{Buettiker1988}
M. B\"{u}ttiker, IBM J. Res. Dev. {\bf 32},  317  (1988).

\bibitem{Kravtsov1993}
V.~E. Kravtsov and V.~I. Yudson, Phys. Rev. Lett. {\bf 70},  210  (1993).

\bibitem{Yudson2003}
V.~I. Yudson and V.~E. Kravtsov, Phys. Rev. B {\bf 67},  155310  (2003).

\bibitem{Buks1996}
E. Buks {\it et~al.}, Phys. Rev. Lett. {\bf 77},  4664  (1996).

\bibitem{Kobayashi2003}
K. Kobayashi, H. Aikawa, S. Katsumoto, and Y. Iye, Phys. Rev. B {\bf 68},
  235304  (2003).

\bibitem{Amaha2008}
S. Amaha {\it et~al.}, Appl. Phys. Lett. {\bf 92},  202109  (2008).

\end{thebibliography}

\end{document}